\documentclass[twocolumn]{aa}
\usepackage{color}
\usepackage{graphicx}
\usepackage{amsmath}	
\usepackage{amssymb}	
\usepackage{mathtools}
\usepackage{txfonts}
\usepackage{booktabs}
\usepackage[T1]{fontenc}
\usepackage{lmodern}
\usepackage{natbib,twoopt}
\usepackage{orcidlink}

\usepackage{float}
\usepackage{subfig}
\usepackage{hyperref}

\usepackage{lscape}

\newcommand{\civ}{C\,\textsc{iv}}

\begin{document}

\title{A measure of cosmological distance using the \civ\ Baldwin effect in quasars}

\author{Long Huang\orcidlink{0000-0003-4545-7066} \inst{1,2}\thanks{\email{huanglong20122021@163.com}}, Hui Wang\inst{3}, Zhifu Gao\inst{4}, Xiangyun Zeng\inst{5}, Zhangyong Chang\inst{1,2}}

 \institute{College of Science, Jiujiang University, Jiujiang 332000, People's Republic of China
     \and
     Key Laboratory of Functional Microscale Materials in Jiangxi Province, Jiujiang 332000, People's Republic of China
\and
School of Physics and Astronomy, China West Normal University, Nanchong 637000, People's Republic of China
\and
Xinjiang Astronomical Observatory, Chinese Academy of Sciences, Urumqi 830011, People's Republic of China
\and
College of Science, China Three Gorges University, Yichang 443000, People's Republic of China}

\titlerunning{A measure of cosmological distance using the \civ\ Baldwin effect in quasars}
  \authorrunning{Long Huang et al.}

\abstract {
We use the anticorrelation between the equivalent width (EW) of the C\,\textsc{iv} 1549 {\AA} emission line and the continuum luminosity in the quasars rest frame (Baldwin effect) to measure their luminosity distance as well as estimate cosmological parameters. We obtain a sample of 471 Type I quasars with the UV-optical spectra and EW (C\,\textsc{iv}) measurements in the redshift range of $2.3< z< 7.1$ including 25 objects at $5 < z < 7.1$, which can be used to investigate the C\,\textsc{iv} Baldwin effect and determine cosmological luminosity distance. The relation $EW(C\,\textsc{iv}) \propto {(\lambda {L_\lambda })^\gamma }$ can be applied to check the inverse correlation between the C\,\textsc{iv} EW and ${L_\lambda }$ of quasars and give their distance, and the data suggest that the EW of C\,\textsc{iv} is inversely correlated with continuum monochromatic luminosities. On the other hand, we also consider dividing the Type I quasar sample into various redshift bins, which can be used to check if the C\,\textsc{iv} EW-luminosity relation depends on the redshift. Finally, we apply a combination of Type I quasars and Type Ia supernovae
(SNIa) of the Pantheon sample to test the property of dark energy concerning whether or not its density deviates from the constant, and give the statistical results.}

\keywords{Quasars; quasars: emission lines; Cosmology; Dark energy}

\maketitle

\section{Introduction}

A wide variety of emission line strengths and velocity widths are the important spectral features of active galactic nuclei (AGNs) and quasars, which can be used to classify these objects and investigate the correlation between the equivalent widths (EWs) of emission lines and continuum luminosities in the UV-optical band. The full width at half maximum (FWHM) of the emission lines often involves the orientation relative to the line of sight \citep{Shen2014}. Broad lines are defined as having $FWHM \approx 1000 - 15,000{\kern 1pt} {\kern 1pt} km{\kern 1pt} {\kern 1pt} {s^{ - 1}}$ and narrow lines as $FWHM \approx 200 - 2000{\kern 1pt} {\kern 1pt} km{\kern 1pt} {\kern 1pt} {s^{ - 1}}$ \citep{Sulentic2000}. On this basis, AGNs and quasars can be categorized by whether they have broad emission lines (Type I), only narrow lines (Type II), or no lines except when a  variable continuum is in a low phase (Blazars) \citep{Urry1995, Sulentic2000}. In addition, other classification methods of quasars can be based on the ratio of monochromatic luminosities. Radio-loud quasars satisfy $logR >1$ and radio-quiet quasars with $logR{\rm{ }} \le 1$, where R is the ratio of monochromatic luminosities (with units of ${\rm{erg}} \cdot {s^{ - 1}} \cdot H{z^{ - 1}}$) measured at (rest-frame) 5GHz and 2500 {\AA} \citep{Strittmatter1980, Kellermann1989, Stocke1992, Kellermann1994}.

On the other hand,  the correlation between the equivalent width (EW) of the C\,\textsc{iv} 1549 {\AA} emission line and the continuum luminosity in the quasars rest frame was investigated by Baldwin \citep{Baldwin1977}. The data suggested that the C\,\textsc{iv}  EW anticorrelates with continuum monochromatic luminosities based on a sample of 20 quasars in the redshift range $1.24 \le z \le 3.53$. This has become known as the Baldwin effect (hereafter BEff). Subsequently, more spectroscopic data of AGNs and quasars are used to verify the BEff, and it exists not only for C\,\textsc{iv} but also for many other UV-optical emission lines such as $Ly\alpha{\kern 1pt}{\kern 1pt} \lambda {\kern 1pt} 1215.7$, $ C\,\textsc{iii}]{\kern 1pt} {\kern 1pt}\lambda {\kern 1pt} 1908.7$, ${\kern 1pt} {\kern 1pt} Mg\,\textsc{ii}{\kern 1pt} {\kern 1pt} {\kern 1pt} {\kern 1pt} {\kern 1pt} \lambda {\kern 1pt} {\kern 1pt} 2800.3$ \citep{Kinney1990, Netzer1992, Croom2002}. Although the physical reason for the BEff remains unknown, there are several explanations that try to account for the UV-optical BEff.

One promising explanation is the softening of the spectral energy distribution (SED) that the soft X-ray continuum between 0.1 and 1kev in high-luminosity quasars is weaker than that in low-luminosity AGNs, which determines the heating rate and the excitation of various collisionally excited lines \citep{Netzer1992, Zheng1995, Dietrich2002}. It is an important clue for the physical cause of the UV-optical BEff. Other underlying physical causes of the BEff involve the black hole mass \citep{Xu2008, Chang2021},  the Eddington ratio $L/{L_{E{\rm{dd}}}}$ \citep{Baskin2004, Xu2008, Dong2009, Nikolajuk2012, Shemmer2015}, and the luminosity dependence of metallicity \citep{Dietrich2002, Sulentic2007}.

In this paper we introduce the source of data used in Section \ref{Sec:2}, including the EW of the C\,\textsc{iv} 1549 {\AA} emission line, and the continuum luminosity at 2500 {\AA} of 471 Type I quasars. In Section \ref{Sec:3} we employ the nonlinear relation $EW(C\,\textsc{iv}) \propto {(\lambda {L_{2500{\kern 1pt}{\AA} }}  )^\gamma }$ to check the correlation between the C\,\textsc{iv} EW and continuum luminosity of Type I quasars, and give their cosmological luminosity distance. In Section \ref{Sec:4} we consider dividing the Type I quasar sample into various redshift bins, which can be used to check if the C\,\textsc{iv} EW-luminosity relation depends on the redshift. In Section \ref{Sec:5}, we apply a combination of Type I quasars and SNla Pantheon to reconstruct the dark energy equation of state $w(z)$, which can be used to test the nature of dark energy concerning whether or not its density deviates from the constant. In Section \ref{Sec:6}, we summarize the paper.

\section[ Data used]{ Data used} \label{Sec:2}

Modern optical instruments and surveys including the Sloan Digital Sky Survey(SDSS) \citep{Lyke2020, Ahumada2020}, the Hubble Space Telescope (HST) \citep{Tacconi2018}, and the International Ultraviolet Explorer (IUE) provide the UV-optical spectra for a large amount of quasars \citep{Kondo1989}, which can be used to investigate the C\,\textsc{iv} BEff. Alam et al. (2015) presented the Data Release 12 Quasar catalog (DR12) data gathered by SDSS-III from 2008 August to 2014 June, which includes the spectra of 294,512 quasars \citep{Alam2015}. Their emission line fluxes or EWs and widths can be measured by different techniques, and sometimes different results are obtained for the same set of data \citep{Berk2001, Shen2008, Shen2011, Shen2012, Paris2011, Paris2012}.

We introduce two main calculation methods for the flux and width of emission lines. The first is to use two Gaussian functions to fit the emission-line spectrum given after subtracting the power-law continuum$({f_\nu } \propto {\nu ^{{\alpha _\nu }}})$. The second method is to employ the principal component analysis (PCA) method to estimate continuum spectrum; this employs the assumption that the covariance matrix of the entire sample can be considered as the covariance matrix of the single sample. Using PCA to fit the continuum and emission line avoids the need to assume a line profile in a region of the spectrum affected by sky subtraction.

\begin{figure}
\subfloat
{\begin{minipage}[b]{0.5\textwidth}
\includegraphics[width=\linewidth,scale=1]{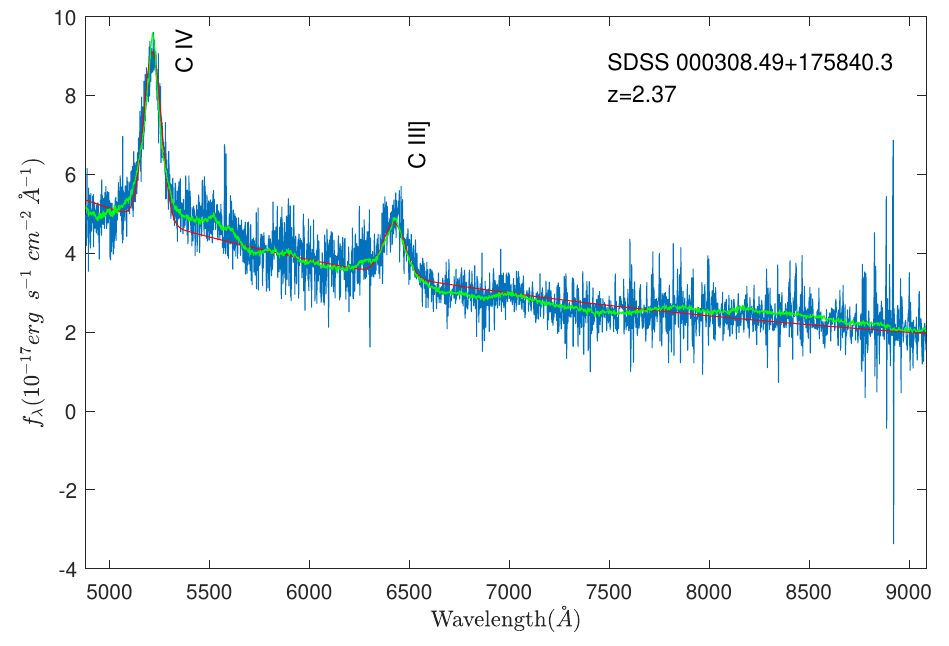}
\end{minipage}
}

\subfloat
{\begin{minipage}[b]{0.5\textwidth}
\includegraphics[width=\linewidth,scale=1]{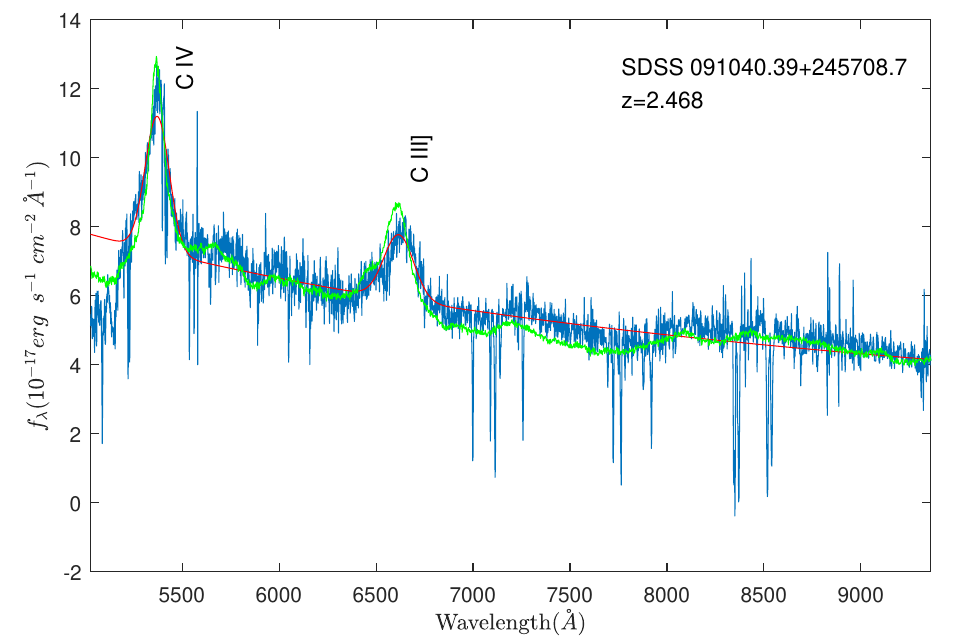}
\end{minipage}
}

\caption{Illustration of two methods used to fit continuum and emission lines.  The red line is a fit of the power-law continuum and two Gaussians functions for $C\,\textsc{iv}$ and $ C\,\textsc{iii}]$  emission lines; the green line is the PCA estimate of the continuum.}
\label{fig:1.1}
\end{figure}

\begin{figure}
\subfloat
{\begin{minipage}[b]{0.5\textwidth}
\includegraphics[width=\linewidth,scale=1]{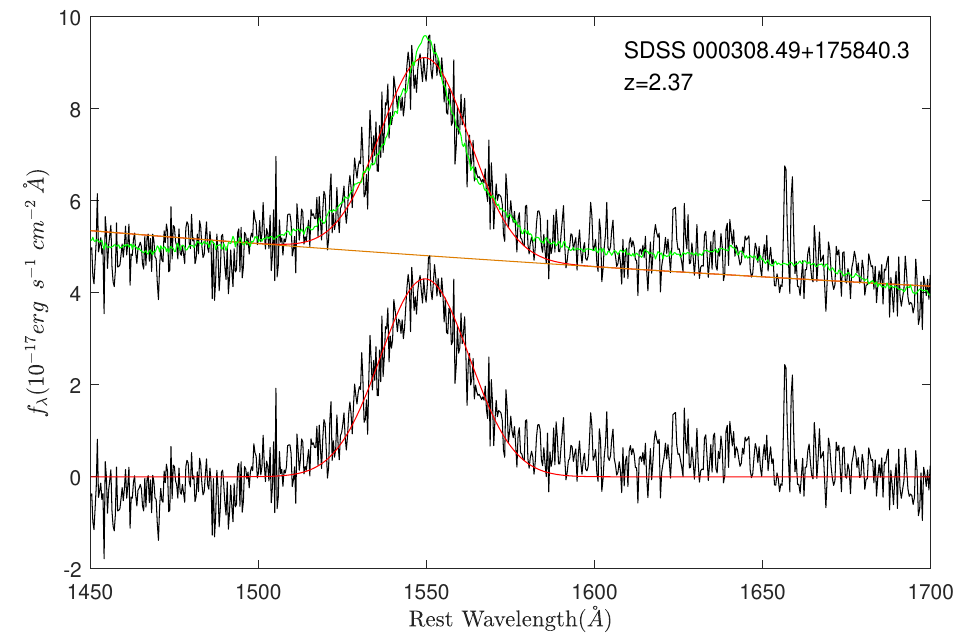}
\end{minipage}
}

\subfloat
{\begin{minipage}[b]{0.5\textwidth}
\includegraphics[width=\linewidth,scale=1]{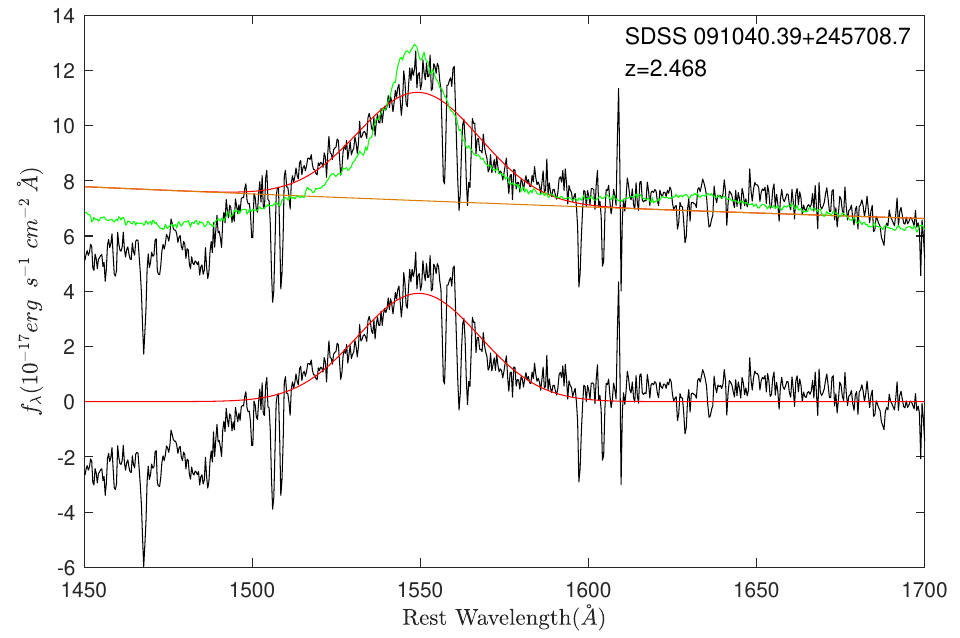}
\end{minipage}
}

\caption{ Two examples of continuum and $C\,\textsc{iv}$ emission line fit. In each panel the
upper black solid line is the original and the lower black solid line is continuum subtracted; the red and green solid curves are the fits to the spectra by Gaussian fit and PCA method; the orange lines show the fitted power-law continuum. }
\label{fig:1.11}
\end{figure}

 We use two Gaussian functions to fit $C\,\textsc{iv}$ and $ C\,\textsc{iii}]$ emission lines with the fitting window 1450-1700 {\AA} and 1800-2000 {\AA} in the rest frame, and the weak emission line $ He\,\textsc{ii}$ is not taken into account. The width and amplitude are independent parameters, but the two Gaussians are bound to have the same emission redshift. Figure \ref{fig:1.11} shows the corresponding C\,\textsc{iv} emission line fit from a two-Gaussian fit. Meanwhile, the PCA method is also applied to fit the spectra. Examples of fitting results for the power-law and Gaussian method and the PCA method are presented in Fig. \ref{fig:1.1}. The C\,\textsc{iv} EW from the two-Gaussian fit is larger than the measurement from the PCA method, and the power-law and Gaussian method gives simpler results to the PCA method, so we consider using the EW of emission lines obtained by the PCA method to study $C\,\textsc{iv}$ Beff. \citet{Paris2017} provided the various measured quantities of  297 301 quasars from DR12 based on the result of a PCA of the spectra. Therefore, we filter samples from their released data to check the correlation between the C\,\textsc{iv} EW and the continuum luminosity for quasars and measure their cosmological luminosity distance.

\begin{figure*}
\begin{center}
\begin{minipage}[t]{0.49\textwidth}
\includegraphics[width=\linewidth,scale=1.00]{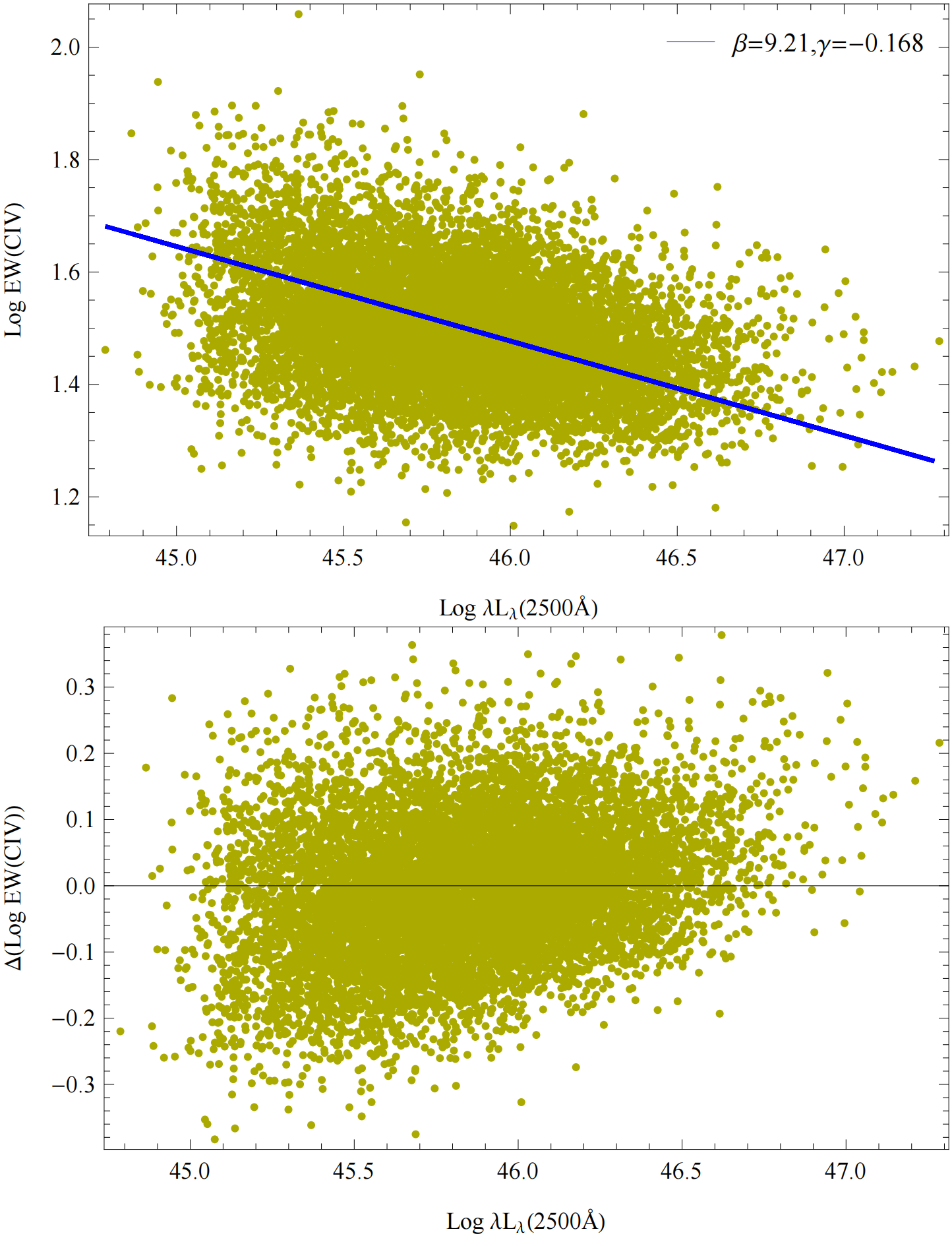}
\caption{Plot of $\log EW(C\,\textsc{iv}) {\kern 1pt}vs. {\kern 1pt}{L_{UV}}$  (upper panel), $\Delta (\log EW(C\,\textsc{iv})) {\kern 1pt}vs. {\kern 1pt}{L_{UV}}$ (lower panel) for Type I quasars. The blue line in the upper panel is the theoretical values of $\log EW(C\,\textsc{iv})$ from Eq. (\ref{eq1}) with the best fitting values of $\beta$ and $\gamma$. $\Delta (\log EW(C\,\textsc{iv}))$ represents the residuals from the subtraction of the observational data and theoretical values. In the lower panel the relationship between $\Delta (\log EW(C\,\textsc{iv}))$ and ${L_{UV}}$ indicates the luminosity dependence of the BEff slope.}
\label{fig:1.2}
\end{minipage}
\begin{minipage}[t]{0.49\textwidth}
\includegraphics[width=\linewidth,scale=1.00]{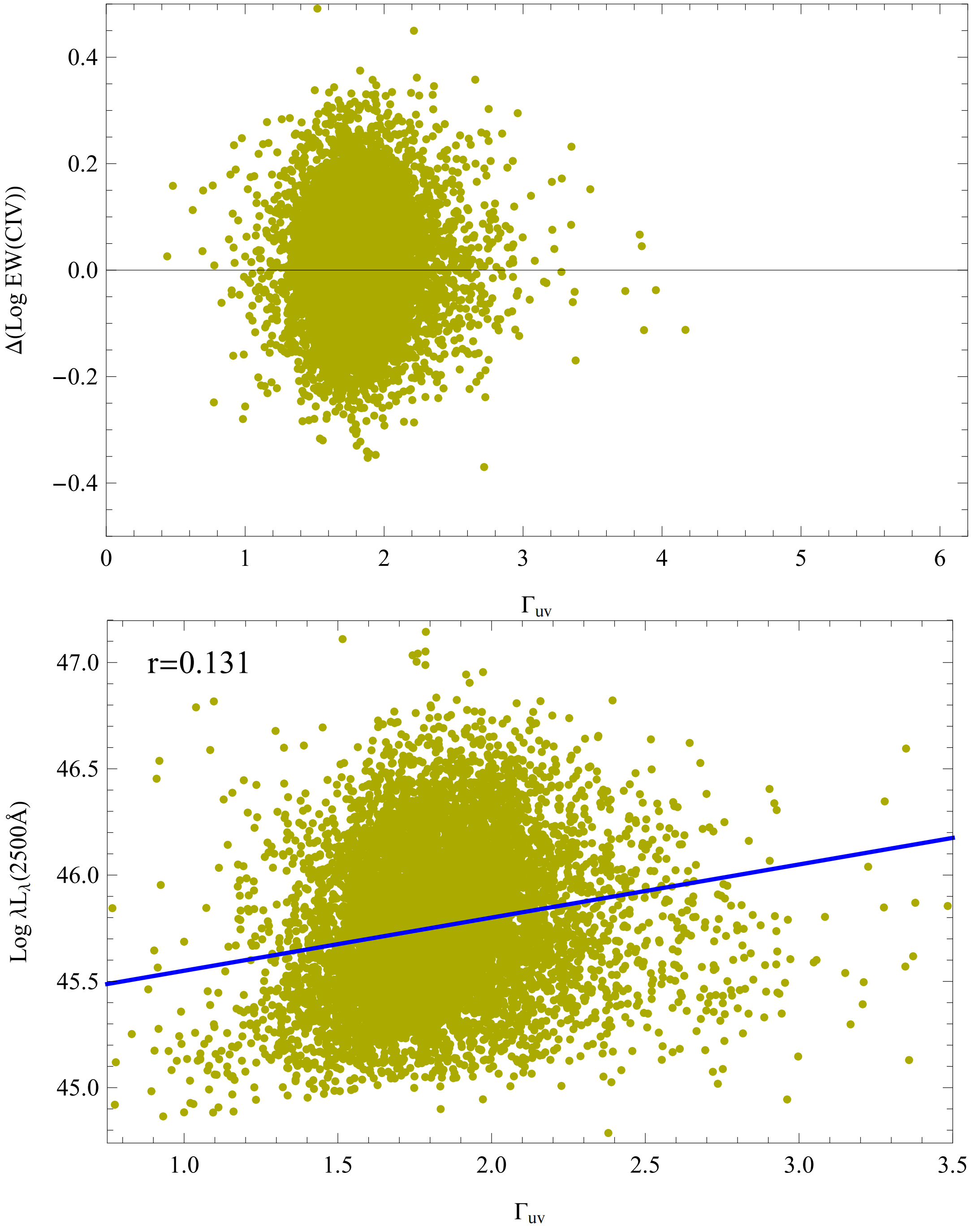}
\caption{Plot of $\Delta (\log EW(C\,\textsc{iv})){\kern 1pt}vs. {\kern 1pt} {\Gamma _{UV}} $ (upper panel) and ${L_{UV}} {\kern 1pt}vs. {\kern 1pt} {\Gamma _{UV}}$ (lower panel) for Type I quasars with $z \le 3$, where ${\Gamma _{UV}}$ is the UV-optical power-law index, and $r$ represents the correlation coefficient.}
\label{fig:1.21}
\end{minipage}
\end{center}
\end{figure*}

\section{Parameter constraints from Type I quasars}\label{Sec:3}
\subsection{Insights from scatter plots}

The linear formula is usually used to investigate the correlation between the C\,\textsc{iv} EW and $\lambda {L_\lambda }$ for quasars, which can be written as

 \begin{equation}\label{eq1}
\log EW(C\,\textsc{iv}) = \beta  + \gamma \log (\lambda {L_\lambda }(2500 {\AA} )).
\end{equation}

This equation is equivalent to the relation $EW(C\,\textsc{iv}) \propto {(\lambda {L_{2500{\kern 1pt}{\AA} }}  )^\gamma }$. We fitted equation (\ref{eq1}) to 8630 Type I quasars and obtained the residual $\Delta (\log EW(C\,\textsc{iv}))$ from the statistical values of $\beta$ and $\gamma$; the luminosities ${L_\lambda }(2500{\AA})$ were obtained from the measured fluxes assuming Lambda cold dark matter ($\Lambda CDM$) cosmology $({\Omega _m} = 0.3,{\kern 1pt} {\kern 1pt} {H_0} = 70{\kern 1pt} km{\kern 1pt} {\kern 1pt} {s^{ - 1}}{\kern 1pt} Mp{c^{ - 1}})$.  The $\log EW(C\,\textsc{iv}) - {L_{UV}}$ plot of Type I quasars are shown in the upper panel of Fig. \ref{fig:1.2}; the lower panel of Fig. \ref{fig:1.2} illustrates  $\Delta (\log EW(C\,\textsc{iv}))$ against ${L_{UV}}$, which implies that the BEff slope $\gamma$ is dependent on luminosity. We also analyzed the $\Delta (\log EW(C\,\textsc{iv}))- {\Gamma _{UV}}$ relation and investigate the correlation between the luminosity ${L_{UV}}$ and the UV-optical power-law index ${\Gamma _{UV}}$, ${\Gamma _{UV}}$ can be obtained from a fit of ${f_\nu } \propto {\nu ^{ - ({\Gamma _{UV}} - 1)}}$ to $u,g,r,i$ and $z$ band. The index ${\Gamma _{UV}}$ of Type I quasars with $z>3$ seem to be inapplicable to study a corresponding correlation and needs to be excluded. The residual $\Delta (\log EW(C\,\textsc{iv}))$ against ${\Gamma _{UV}}$ for Type I quasars with $z\le3$ is shown in the upper panel of Fig. \ref{fig:1.21}; it implies that $\Delta (\log EW(C\,\textsc{iv}))$ has no dependence on ${\Gamma _{UV}}$. The luminosity ${L_{UV}}$ against ${\Gamma _{UV}}$ is provided in the lower panel of Fig. \ref{fig:1.21}; their correlation coefficient is $r =  0.131$, which indicates that the luminosity ${L_{UV}}$  might be weakly correlated with the UV-optical power-law index ${\Gamma _{UV}}$.

\subsection{Filter data for measuring luminosity distance}

As can be seen in the lower panel of Fig. \ref{fig:1.2}, the luminosity dependence of the BEff slope is indicated for a large number of samples, so we fitted our data points with $45.8 \le \log (\lambda {L_\lambda }(2500{\AA})) \le 46.5$ to avoid the luminosity dependence for $\gamma$. Meanwhile, in order to reduce the dispersion, we select the sample at $z \le 3$ by $\sigma (EW)/EW(C\,\textsc{iv}) < 0.02$, $\sigma (EW)/EW(C\,\textsc{iv}) < 0.023$ for $3 < z \le 5$, $\sigma (EW)$ is the uncertainty for EW (C\,\textsc{iv})), and the higher redshift $z > 5$ with $\sigma (EW)/EW(C\,\textsc{iv}) < 0.06$, which ensures that the numbers of objects are close in different redshift bins. We then obtain a sample of 471 Type I quasars($2.3< z <7.1$). We also match the sample to the latest FIRST catalog and The NRAO VLA sky survey (NVSS) data using a $2''$ matching radius \citep{Helfand2015, Condon1998}, only about 20 objects have FIRST and NVSS counterparts, but all of them satisfy $\log R > 1$, these radio-loud sources and other quasars compose a total sample of 471 Type I quasars. We can use this sample to investigate BEff and calculate the luminosity distance.

\subsection{Parametric formula for C\,\textsc{iv} EW and the continuum flux}

 Using relation $L = 4\pi {D_L}^2f$ in equation (\ref{eq1}), we get
\begin{equation}\label{eq2}
\begin{array}{*{20}{l}}
{\log EW(C\,\textsc{iv}) = \Phi (\lambda {f_\lambda }(2500{\kern 1pt} {\kern 1pt} {\AA}),{D_L})}\\
{{\kern 1pt} {\kern 1pt} {\kern 1pt} {\kern 1pt} {\kern 1pt} {\kern 1pt} {\kern 1pt} {\kern 1pt} {\kern 1pt} {\kern 1pt} {\kern 1pt} {\kern 1pt} {\kern 1pt} {\kern 1pt} {\kern 1pt} {\kern 1pt} {\kern 1pt} {\kern 1pt} {\kern 1pt} {\kern 1pt} {\kern 1pt} {\kern 1pt} {\kern 1pt} {\kern 1pt} {\kern 1pt} {\kern 1pt} {\kern 1pt} {\kern 1pt} {\kern 1pt} {\kern 1pt}  = \beta  + \gamma \log (\lambda {f_\lambda }(2500{\kern 1pt} {\kern 1pt} {\AA})) + \gamma \log (4\pi {D_L}^2)},\\
\end{array}
\end{equation}
where ${f_\lambda }(2500{\kern 1pt} {\kern 1pt} {\AA})$ is the flux measured at (rest-frame) $2500 {\AA}$; EW(C\,\textsc{iv}) is the equivalent width of the C\,\textsc{iv} 1549 {\AA} emission line; and $D_L$ is the luminosity distance, which depends on the redshift $z$. Thus, equation (\ref{eq2}) can be used to check the C\,\textsc{iv} EW-luminosity correlation for Type I quasars and determine their cosmological luminosity distance.

We fit the C\,\textsc{iv} EW-luminosity relation to Type I quasars by minimizing a likelihood function consisting of a modified ${\chi ^2}$ function based on a Markov chain Monte Carlo (MCMC) function, allowing for an intrinsic dispersion $\delta$
 \begin{equation}\label{eq3}
\begin{array}{l}
 - 2\ln L = \sum\limits_{i = 1}^N {\left\{ {\frac{{{{[\log EW{{(C\,\textsc{iv})}_i} - \Phi {{(\lambda {f_\lambda }(2500{\kern 1pt} {\kern 1pt} {\AA}),{D_L})}_i}]}^2}}}{{s_i^2}}} \right\}} \\
{\kern 1pt} {\kern 1pt} {\kern 1pt} {\kern 1pt} {\kern 1pt} {\kern 1pt} {\kern 1pt} {\kern 1pt} {\kern 1pt} {\kern 1pt} {\kern 1pt} {\kern 1pt} {\kern 1pt} {\kern 1pt} {\kern 1pt} {\kern 1pt} {\kern 1pt} {\kern 1pt} {\kern 1pt} {\kern 1pt} {\kern 1pt} {\kern 1pt} {\kern 1pt} {\kern 1pt} {\kern 1pt} {\kern 1pt} {\kern 1pt} {\kern 1pt} {\kern 1pt} {\kern 1pt} {\kern 1pt}  + \sum\limits_{i = 1}^N {\ln (2\pi s_i^2)},
\end{array}
\end{equation}
where ${\Phi {{(\lambda {f_\lambda }(2500{\kern 1pt} {\kern 1pt} {\AA}),{D_L})}_i}}$ is given by equation (\ref{eq2}) and ${s_i}^2 = \sigma _i^2(\log {\kern 1pt} EW (C\,\textsc{iv})) + {\gamma ^2} \cdot \sigma _i^2(\log {\kern 1pt} {\kern 1pt} (\lambda {f_\lambda }(2500{\AA}))) + {\delta ^2}$;  $\sigma _i(\log {\kern 1pt} EW)$ and $\sigma _i(\log {\kern 1pt} {\kern 1pt} (\lambda {f_\lambda }(2500{\AA})))$ indicate the statistical errors for $\log EW(C\,\textsc{iv})$ and $\log (\lambda{f_\lambda }(2500 {\AA}))$; and $\delta$ is the intrinsic dispersion \citep{Kim2011,Risaliti2015}, which can be fitted as a free parameter.

We do not consider the large positive and negative space curvature as they have not been obviously observed by observational data, and we approximately adopt a curvature of ${\Omega _k} = 0$;  a prior cosmological constant $\Lambda CDM$ model is assumed, then ${\Omega _\Lambda }{\rm{ = }}1{\rm{ - }}{\Omega _m}$ when ${\Omega _R} \ll {\Omega _m}$. In this case the free parameters are $\beta$, $\gamma$ and the intrinsic dispersion $\delta$, and the cosmological parameters ${\Omega _m}$. We note that the Hubble constant ${H_0}$ is absorbed into the parameter $\beta$ when fitting equation (\ref{eq2}), without an independent determination of this parameter, so we fix ${H_0} = 70{\kern 1pt} {\kern 1pt} km{\kern 1pt} {\kern 1pt} {\kern 1pt} {s^{ - 1}}{\kern 1pt} Mp{c^{ - 1}}$ \citep{Reid2019, Aghanim2020}.

Meanwhile, we measure the distance modulus for Type I quasars based on the C\,\textsc{iv} EW-luminosity relation. On the other hand, other methods for quasars also can be applied to measure their luminosity distance \citep{La2014, Risaliti2015, Martinez2019, Dultzin2020}. Equation (\ref{eq2}) gives the distance modulus as
 \begin{equation}\label{eq4}
{\rm{DM  = }}\frac{{5[\log EW(C\,\textsc{iv}) - \gamma \log (\lambda {f_\lambda }(2500{\kern 1pt} {\kern 1pt} {\AA})) - \beta ']}}{{2\gamma }}+25,
\end{equation}
where $\beta ' = \beta  + \gamma \log {\kern 1pt} (4\pi )$.  The error is
 \begin{equation}\label{eq5}
{\sigma _{DM}} = DM\sqrt {{{(\frac{{{\sigma _f}}}{f})}^2} + {{(\frac{{{\sigma _\gamma }}}{\gamma })}^2}} ,
\end{equation}
where $f=\log EW(C\,\textsc{iv}) - \gamma \log (\lambda {f_\lambda }(2500{\kern 1pt} {\kern 1pt} {\AA})) - \beta '$, and ${\sigma _f}^2 = {\sigma ^2}(\log {\kern 1pt} EW (C\,\textsc{iv})) + {\gamma ^2} \cdot {\sigma ^2}(\log {\kern 1pt} {\kern 1pt} (\lambda {f_\lambda }(2500{\AA})))+{\sigma _{\beta '}}^2$. From equation (\ref{eq5}),  the uncertainty of the slope of the BEff $\gamma$ obviously influences the error of distance modulus for Type I quasars.

\begin{table*}
\centering
\caption{Properties of the 471 Type I quasars. $DM$ is the distance modulus from a fit of the relation $EW(C\,\textsc{iv} ) \propto {(\lambda {L_{\lambda }})^\gamma }$ with $\Lambda CDM$ model; ${\sigma _{DM}}$ is the error. Only five of the objects are listed.}
\label{tab:3}
\vspace{0.3cm}
\begin{tabular}{cccccccc}
\hline
SDSS name &$z $&${m_r}$&$EW(C\,\textsc{iv})$&$\log {f_\nu  }(2500{\AA}) $&${\Gamma _{UV}}$&$DM$&${\sigma _{DM}}$\\
&&mag&${\AA}$&$erg{\kern 1pt} {\kern 1pt} {s^{ - 1}}c{m^{ - 2}}H{z^{ - 1}}$&\\
\hline

143112.39+093915.4	&7.011	&21.094$\pm$0.047	&26.56$\pm$0.97&	-24.536$\pm$0.019&	2.964	&49.389	&1.956\\
112310.06+134622.5	&6.038	&21.06$\pm$0.038	&27.49$\pm$1.13	&-24.505$\pm$0.015	&4.383	&49.077	&1.949\\
115132.69+550317.3	&5.338	&21.166$\pm$0.054	&28.96$\pm$1.68	&-24.534$\pm$0.022	&3.823	&48.789	&1.959\\
125718.02+374729.9	&4.745	&20.757$\pm$0.037	&29.16$\pm$0.48	&-24.358$\pm$0.015	&5.042	&48.302	&1.904\\
131808.44+215437.0	&4.258	&20$\pm$0.032	     &28.06$\pm$0.58	&-24.043$\pm$0.013	&4.245	&47.78	&1.886\\

\hline

\end{tabular}

\end{table*}

\begin{figure*}
\centering
\includegraphics[width=500pt]{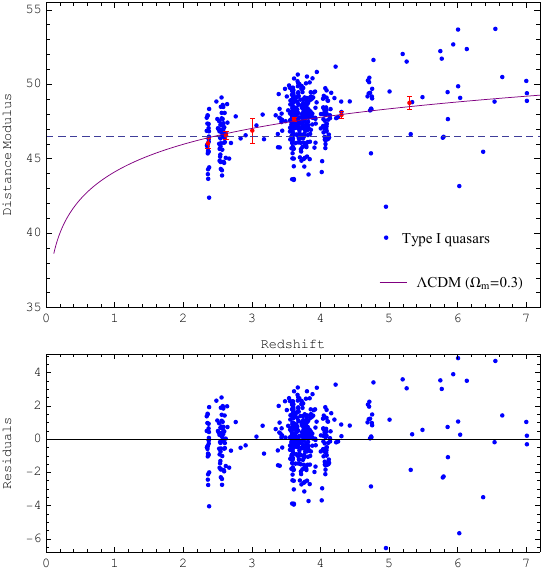}
\caption{Type I quasar distance modulus (blue points) from a fit of Equation (\ref{eq2}) when assuming $\Lambda CDM$ cosmology. The red points and bars (upper panel) are distance modulus averages and the standard deviations of the mean in small redshift bins. The purple line shows a flat $\Lambda CDM$ model fit with ${\Omega _m} = 0.3$; the dotted line is the reference distance modulus and its value is 46.5. The lower panel shows the residuals of the distance modulus at different redshifts. }
\label{fig:2}
\end{figure*}

\begin{figure}
\centering
\includegraphics[width=\linewidth,scale=1.00]{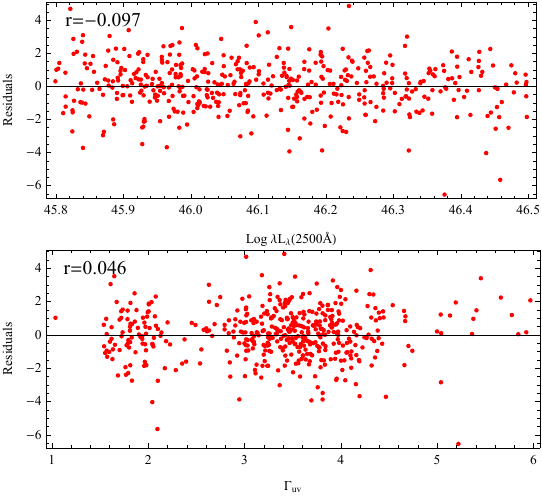}
\caption{Plot of residuals of distance modulus against luminosity ${L_{UV}}$ or UV-optical power-law index ${\Gamma _{UV}}$ for the selected sample. This is used to measure the luminosity distance, and $r$ is the correlation coefficient.}
\label{fig:2.1}
\end{figure}

\subsection{Fitting result for the relation of C\,\textsc{iv} EW to the continuum flux}

We adopt the maximum likelihood function (equation (\ref{eq3})) based on MCMC to constrain the parameters; the fitting results are shown in Table \ref{tab:1}, and the slope of the BEff is $\gamma  =  - 0.164 \pm 0.006$, which suggests that the EW of C\,\textsc{iv} is inversely correlated with continuum monochromatic luminosities. It is consistent with the result by \citet{Bian2012}.

We obtain the distance modulus of Type I quasars by substituting the statistical average values of $\beta$ and $\gamma$ into equation (\ref{eq4}), which are shown in the upper panel of Fig. \ref{fig:2}, including their averages in small redshift bins. Meanwhile the distance modulus and properties of the 471 Type I quasars are listed in Table \ref{tab:3}. The lower panel of Fig. \ref{fig:2} shows the plot of the residuals of the distance modulus against redshift; the residuals are from measuring the distance modulus for Type I quasars and $\Lambda CDM$ cosmology $({\Omega _m} = 0.3)$. There could be several reasons for the large scatter in the luminosity distance, including observational error, and intrinsic variation of the BEff \citep{Shields2006, Dietrich2002}. Figure \ref{fig:2.1} illustrates the diagram of the residuals against the luminosity ${L_{UV}}$ or the UV-optical power-law index ${\Gamma _{UV}}$. The correlation coefficient for the residuals $\Delta (\log EW(C\,\textsc{iv}))$ and ${L_{UV}}$ is $r =  - 0.097$, which represents that $\Delta (\log EW(C\,\textsc{iv}))$ is not correlated with the luminosity ${L_{UV}}$. The residuals $\Delta (\log EW(C\,\textsc{iv}))$ is also not relevant with ${\Gamma _{UV}}$; their correlation coefficient is $r =  0.046$, which implies that the BEff slope has no dependence on ${L_{UV}}$ or ${\Gamma _{UV}}$ in the final sample.

We also consider the relevance of the  $C\,\textsc{iv}$ EW dependence on Eddington ratio $L/{L_{E{\rm{dd}}}}$ and use the relation $\log EW(C\,\textsc{iv}) = \alpha  + \beta \log {\kern 1pt} L/{L_{Edd}}$ to measure luminosity distance for quasars \citep{Baskin2004, Bian2012, Ge2016}. The virial black hole (BH) Masses ${M_{BH}}$ or Eddington ratio $L/{L_{E{\rm{dd}}}}$ can be estimated from $C\,\textsc{iv}$ emission lines \citep{Shen2008, Shen2011} by formula $\log ({M_{BH}}/{M_ \odot }) = a + b{\kern 1pt} {\kern 1pt} \log (\lambda {\kern 1pt} {\kern 1pt} {L_\lambda }/({10^{44}}{\kern 1pt} {\kern 1pt} erg{\kern 1pt} {\kern 1pt} {s^{ - 1}})) + 2{\kern 1pt} {\kern 1pt} \log {\kern 1pt} {\kern 1pt} (FWHM/(km{\kern 1pt} {\kern 1pt} {s^{ - 1}}))$, which is derived from the so-called virial black hole mass estimate ${M_{BH}} = {G^{ - 1}}{R_{BLR}}V_{BLR}^2$ and ${R_{BLR}} - \lambda {L_\lambda }$ relation, and $a = 0.66,{\kern 1pt} {\kern 1pt} {\kern 1pt} b = 0.53$ for $C\,\textsc{iv}$ estimators \citep{Mclure2004, Vestergaard2006}. Then we can obtain Eddington ratios $L/{L_{E{\rm{dd}}}}$, where ${L_{E{\rm{dd}}}} = 1.3 \times {10^{38}}({M_{BH}}/{M_ \odot }){\kern 1pt} erg{\kern 1pt} {\kern 1pt} {s^{ - 1}}$ is the Eddington luminosity. We use the $C\,\textsc{iv}$ EW-$L/{L_{E{\rm{dd}}}}$ relation to measure the luminosity distance for Type I quasars. Nonetheless, there are even greater errors in the luminosity distance than the results from the $C\,\textsc{iv}$ EW-luminosity relation $EW(C\,\textsc{iv} ) \propto {(\lambda {L_{\lambda }})^\gamma }$.  Therefore, we only use the distance modulus for Type I quasars obtained from CIV Beff and SNIa Pantheon to test the property of dark energy in Section \ref{Sec:5}.

\begin{table*}
\scriptsize
 \centering
  \caption{Fit results on model parameters for a combination of Type I quasars and SNla}
  \label{tab:2}
  \vspace{0.3cm}
  \begin{tabular}{@{}ccccccccccc@{}}
    \hline

&Parameter&$\beta $&${\gamma} $&$\delta $&${\Omega _m}$&${w_0}$&${w_\alpha }$&${\chi'} _{Total}^2$/$\chi _{Total}^2$/N\\
\hline

    &$Sample $&&&$Quasars (T{\rm{ype}}{\kern 1pt} {\kern 1pt} {\kern 1pt} {\kern 1pt} I)$\\
   $\Lambda CDM$ &Best Fit& 8.84   & -0.16&   0.092 & 0.268 &$-$&$-$ &-902.9/475.9/471\\
   &Mean & 8.89$\pm$0.031   & -0.164$\pm$0.006&   0.103$\pm$0.004 & 0.23$\pm$0.01 &$-$&$-$ &\\
   \hline

   & &&&$SN+Quasars (T{\rm{ype}}{\kern 1pt} {\kern 1pt} {\kern 1pt} {\kern 1pt} I)$\\
   $\Lambda CDM$ &Best Fit& 8.756   & -0.158&   0.092 & 0.273 &$-$&$-$ &134.1/1512.9/1519\\
   &Mean& 8.734$\pm$0.03   & -0.158$\pm$0.006&   0.092$\pm$0.003 & 0.269$\pm$0.006 &$-$&$-$ &\\
${w_0}{w_a}CDM$ &Best Fit& 8.71   & -0.157&    0.092 & 0.308 &-1.13&0.419&130.1/1509/1519\\
&Mean& 8.72$\pm$0.024   & -0.157$\pm$0.006&    0.092$\pm$0.003 & 0.295$\pm$0.017 &-1.07$\pm$0.074&0.108$\pm$0.294&\\
   \hline

\end{tabular}
\label{tab:1}
\end{table*}

\section{Analysis of the relation $EW(\civ ) \propto {(\lambda {L_{\lambda }})^\gamma }$ as a function of $z$}\label{Sec:4}

\begin{figure}
\includegraphics[width=\linewidth,scale=1.00]{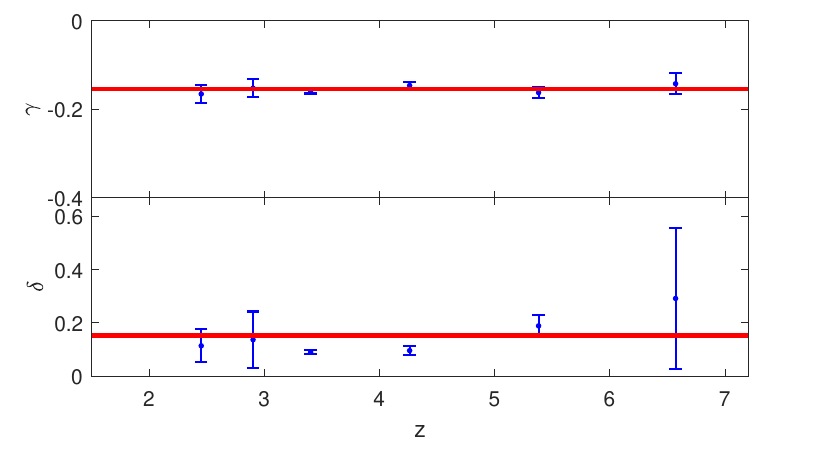}
\caption{C\,\textsc{iv} EW-luminosity correlation in narrow redshift intervals. The blue points are the fit results of ${\gamma}(z),{\kern 1pt} {\kern 1pt} \delta (z)$ at different redshifts. The horizontal lines show their average values.}
\label{fig:3}
\end{figure}

We divide the Type I quasar data into several redshift bins, which can be used to check if relation $EW(C\,\textsc{iv} ) \propto {(\lambda {L_{\lambda }})^\gamma }$ depends on redshift. The redshift bins satisfy $\Delta (1/(1 + z)) = 0.033$. We adopt the parametric model
 \begin{equation}\label{eq6}
\begin{array}{*{20}{l}}
{\log EW(C\,\textsc{iv}) = \beta (z) + \gamma (z)\log (\lambda {f_\lambda }(2500{\kern 1pt} {\kern 1pt} {\AA}))},\\
\end{array}
\end{equation}
where $\gamma (z)$ and the intrinsic dispersion $\delta (z)$ are free parameters, and $\beta (z)$ is obtained from  equation (\ref{eq2}) and can also be a free parameter. We apply segmented Type I quasars data to fit $\gamma (z)$ as well as test whether there is a dependency upon redshift. The fit results of $\gamma (z)$ and ${\kern 1pt} {\kern 1pt} \delta (z)$ at different redshifts are illustrated in Fig. \ref{fig:3}; it is easy to see that their values do not obviously deviate from the average, which shows there is no obvious evidence for any significant redshift evolution. The average values of parameter is $\left\langle \gamma  \right\rangle  =  - 0.154 \pm 0.009.$

\section{The reconstruction of dark energy equation of state w(z)}\label{Sec:5}

Although dark energy can be used to effectively explain the accelerating expansion of the universe and the cosmic microwave background (CMB) anisotropies \citep{Riess1998, Amanullah2010, Betoule2014, Scolnic2018, Conley2010, Hu2002, Ade2016}, the origin and property of its density and pressure remain unclear.

Dark energy can be researched using two main methods. One is to constrain dark energy physical models from observational data and try to explain the physical origin of its density and pressure \citep{Peebles2003, Ratra1988, Li2004, Maziashvili2007, Amendola2000, Gao2017}. Understanding the physical origin of dark energy is important for our universe. It is necessary to determine whether the dark energy is composed of Boson pairs in a vacuum, Fermion pairs, or the Higgs field, and whether it has weak isospin, which may determine whether it can be detected directly in the laboratory. The other method is to study the properties of dark energy, focusing on whether or not its density evolves with time. This can be tested by reconstructing the dark energy equation of state $w(z)$ \citep{Linder2003, Maor2002}, which does not depend on physical models. High-redshift observational data such as quasars can better solve these issues.

 The reconstruction of the equation of state of dark energy $w(z)$ includes parametric and non-parametric methods \citep{Huterer2003, Clarkson2010, Holsclaw2010, Seikel2012, Crittenden2009, Zhao2012}. We employ Type I quasars and Type Ia supernova (SNla) to reconstruct $w(z)$ by the parametric method assuming C\,\textsc{iv} EW-luminosity relation Equation (\ref{eq2}), which can be used to test the nature of dark energy.

For SNIa data, the Pantheon sample contains 1048 SNIa from the Pan-STARRS1 (PS1), the Sloan Digital Sky Survey (SDSS), SNLS, and various low-z and Hubble Space Telescope samples. There are 279 SNIa provided by PS1 \citep{Scolnic2018}, and SDSS presented 335 SNla \citep{Betoule2014,Gunn2006,Sako2018}. The rest of the Pantheon sample are from the ${\rm{CfA1 - 4}}$, CSP, and Hubble Space Telescope (HST) SN surveys \citep{Amanullah2010, Conley2010}. This combined sample of 1048 SNIa is called the Pantheon sample.

The integral formula of the luminosity-redshift relation in flat space can be written as \citep{Linder2003, Sahni2006}
 \begin{equation}\label{eq13}
\begin{array}{l}
{D_L} = \frac{{1 + z}}{{{H_0}}}\int_0^z {d{z'}[{\Omega _m}{{(1 + {z'})}^3}} \\
{\kern 1pt} {\kern 1pt} {\kern 1pt} {\kern 1pt} {\kern 1pt} {\kern 1pt} {\kern 1pt} {\kern 1pt} {\kern 1pt} {\kern 1pt} {\kern 1pt} {\kern 1pt} {\kern 1pt} {\kern 1pt} {\kern 1pt} {\kern 1pt} {\kern 1pt} {\kern 1pt} {\kern 1pt} {\kern 1pt} {\kern 1pt} {\kern 1pt} {\kern 1pt} {\kern 1pt} {\kern 1pt} {\kern 1pt} {\kern 1pt} {\kern 1pt} {\kern 1pt} {\kern 1pt} {\kern 1pt} {\kern 1pt} {\kern 1pt} {\kern 1pt} {\kern 1pt} {\kern 1pt} {\kern 1pt} {\kern 1pt} {\kern 1pt} {\kern 1pt} {\kern 1pt} {\kern 1pt} {\kern 1pt} {\kern 1pt} {\kern 1pt} {\kern 1pt} {\kern 1pt} {\kern 1pt} {\kern 1pt} {\kern 1pt} {\kern 1pt}  + {\Omega _R}{(1 + {z'})^4} + \Omega _{DE}^{(0)}{{\mathop{\rm e}\nolimits} ^{\int_0^{{z'}} {\frac{{1 + w({z^{''}})}}{{1 + {z^{''}}}}d{z^{''}}} }}{]^{ - 1/2}},
\end{array}
\end{equation}
where ${{\Omega _R}}$, ${\Omega _m}$, and ${\Omega _{DE}^{(0)}}$ are the present radiation density, matter density, and dark energy density and satisfies $\Omega _{DE}^{(0)} = 1 - {\Omega _m}$ when ignoring ${{\Omega _R}}$, $w(z)$ is the dark energy equation of state.
We adopt the parametric form for $w(z)$
 \begin{equation}\label{eq13}
w(z) = {w_0} + {w_a}\frac{z}{{1 + z}}.
\end{equation}
and denote it ${w_0}{w_a}CDM$. Therefore dark energy density is
 \begin{equation}\label{eq14}
{\Omega _{DE}}(z) = \Omega _{DE}^{(0)}{(1 + z)^{3(1 + {w_0} + {w_a})}}\exp [ - 3{w_a}z/(1 + z)].
\end{equation}

We constrain the ${{\rm{w}}_0}{{\rm{w}}_a}{\rm{CDM}}$ model parameters for Type I quasars and SNla by minimizing a modified ${\chi'}_{Total}^2$ function. The ${\chi'}_{Total}^2$ is
 \begin{equation}\label{eq15}
{\chi'}_{Total}^2 =  - 2\ln {L^{Quasars}} + \chi _{SN}^2,
\end{equation}
where $ - 2\ln {L^{Quasars}}$ is given by equation (\ref{eq3}), and $\chi _{SN}^2$ can be expressed as
 \begin{equation}\label{eq16}
\chi _{SN}^2 = \Delta {\mu ^T}C_{{\mu _{ob}}}^{ - 1}\Delta \mu ,
\end{equation}
where $\Delta \mu  = \mu  - {\mu _{th}}$; ${C_\mu }$ is the covariance matrix of the distance modulus $\mu $; $\chi _{Total}^2$ function also satisfies $\chi _{Total}^2 = \chi _{Quasar{\rm{s}}}^2 + \chi _{SN}^2$; and $\chi _{Quasar{\rm{s}}}^2 =  - 2\ln {L^{Quasars}} - \sum\limits_{i = 1}^N {\ln (2\pi s_i^2)} $.

\begin{figure}
\includegraphics[width=\linewidth]{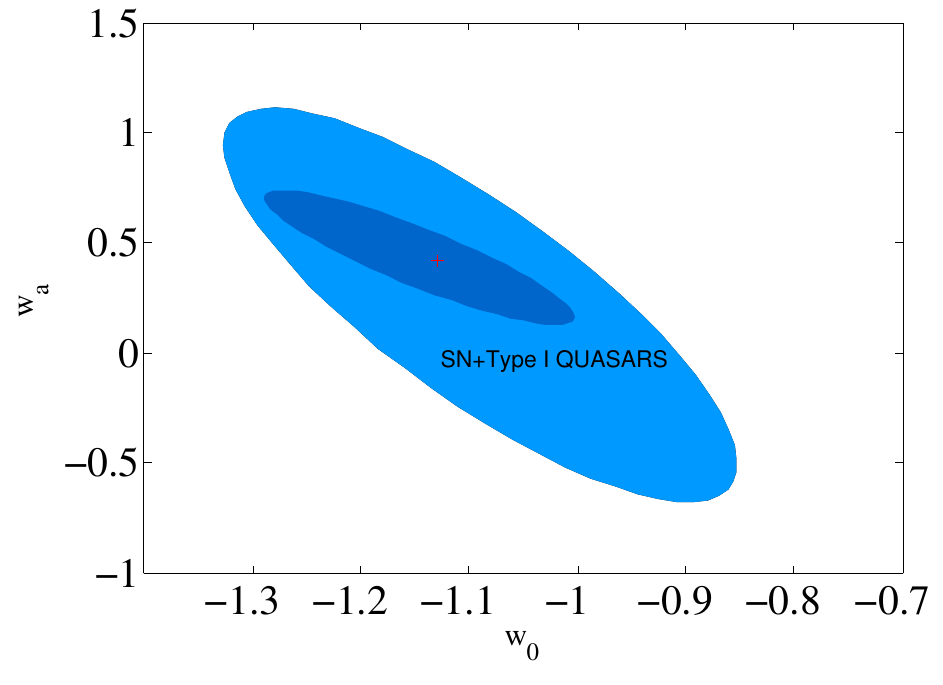}
\caption{Contours at $68\%$ and $95\%$ $(\Delta {\chi ^2} = 1.5,{\kern 1pt} {\kern 1pt} 6)$ for ${w_0}$ and ${w_a}$ from a fit of the relation $EW(C\,\textsc{iv} ) \propto {(\lambda {L_{\lambda }})^\gamma }$ with ${w_0}{w_a}CDM$ model to a combination of SNla and Type I quasars. The plus sign (+) in the corresponding color represents the best fitting values for ${w_0}$, ${w_a}$.}
\label{fig:4}
\end{figure}

We use Type I quasars and SN la to fit the equation (\ref{eq15}) and obtain the statistical results for the cosmological parameters, and their mean and best fit values are listed in Table \ref{tab:2}. When using a combination of Type I quasars and SNla which covers low- and high-redshift data, the results show ${w_0}{w_a}CDM$ has better goodness of fit than $\Lambda CDM$, and ${\chi '}_{Total}^2$ is improved by $-3.9$, which indicates that the $\Lambda CDM$ model is in tension with Type I quasars and SNIa at $\sim 1.5\sigma$. The results are consistent with the values from radio-loud quasars \citep{Huang2022}. Meanwhile, Fig. \ref{fig:4} illustrates the $68\%$ and $95\%$ contours for ${w_0}$ and ${w_a}$ from a combination of SNla and Type I quasars, assuming the C\,\textsc{iv} EW-luminosity relation $EW(C\,\textsc{iv} ) \propto {(\lambda {L_{\lambda }})^\gamma }$.

\section{Summary}\label{Sec:6}

 Although the precise physical reason for the connection between the EW of C IV and continuum luminosity for quasars is not clearly understood, we can still  use a parametric formula to check their correlation. We obtained a sample of 471 Type I quasars with the C\,\textsc{iv} EW of emission lines and continuum luminosity in the UV-optical band including 25 objects at $5 < z < 7.1$, which can be more effective for checking cosmological model and testing the nature of dark energy. First, we adopted the parametric formula to check the correlation between the C\,\textsc{iv} EW and luminosity, and obtained a C\,\textsc{iv} BEff slope $\gamma  =  - 0.164 \pm 0.006$, which implies the EW of C\,\textsc{iv} is inversely correlated with continuum monochromatic luminosity.

Second, We divided the Type I quasars data into several redshift bins and combined a special model to check if there is a redshift evolution of the C\,\textsc{iv} EW-luminosity relation. The fitting results show the slope $\gamma (z)$ approaches the constant, which shows that there is not an obvious redshift evolution for $EW(C\,\textsc{iv} ) \propto {(\lambda {L_{\lambda }})^\gamma }$.

Finally, we used a combination of Type I quasars and SNla to test the property of dark energy by reconstructing the equation of state $w(z)$. The 471 high-redshift Type I quasars included 25 objects at $5 < z < 7.1$, which can be applied to check the cosmological model and test the nature of dark energy more efficiently. The results show the ${w_0}{w_a}CDM$ model is superior to the cosmological constant $\Lambda CDM$ model at $\sim 1.5\sigma $.

In the future, we will select more Type I quasars at high redshift ($z>5$) with the C\,\textsc{iv} 1549 {\AA} emission line and the continuum luminosity from the SDSS quasar catalogs, and hope to obtain $z>7$ quasars from future optical observations, such as the James Webb Space Telescope (JWST) \citep{Naidu2022}. The high-redshift observational data can be better used to reconstruct the equation of state and test the properties of dark energy, which involves whether or not the universe will keep expanding.

\begin{acknowledgements}
This work was supported by the National Natural Science Foundation of China (No.12203029).
\end{acknowledgements}

\vspace{1cm}
\bibliographystyle{aa}
\bibliography{bib}

\begin{thebibliography}{72}
\expandafter\ifx\csname natexlab\endcsname\relax\def\natexlab#1{#1}\fi

\bibitem[{Ade {et~al.}(2016)Ade, Aghanim, Arnaud, Ashdown, Aumont, Baccigalupi,
  Banday, Barreiro, Bartlett, Bartolo, {et~al.}}]{Ade2016}
Ade, P., Aghanim, N., Arnaud, M., {et~al.} 2016, A\&A, 594, A24

\bibitem[{Aghanim {et~al.}(2020)Aghanim, Akrami, Ashdown, Aumont, Baccigalupi,
  Ballardini, Banday, Barreiro, Bartolo, Basak, {et~al.}}]{Aghanim2020}
Aghanim, N., Akrami, Y., Ashdown, M., {et~al.} 2020, A\&A, 641, A6

\bibitem[{Ahumada {et~al.}(2020)Ahumada, Prieto, Almeida, Anders, Anderson,
  Andrews, Anguiano, Arcodia, Armengaud, Aubert, {et~al.}}]{Ahumada2020}
Ahumada, R., Prieto, C.~A., Almeida, A., {et~al.} 2020, APJS, 249, 3

\bibitem[{Alam {et~al.}(2015)Alam, Albareti, Prieto, Anders, Anderson,
  Anderton, Andrews, Armengaud, Aubourg, Bailey, {et~al.}}]{Alam2015}
Alam, S., Albareti, F.~D., Prieto, C.~A., {et~al.} 2015, APJS, 219, 12

\bibitem[{Amanullah {et~al.}(2010)Amanullah, Lidman, Rubin, Aldering, Astier,
  Barbary, Burns, Conley, Dawson, Deustua, {et~al.}}]{Amanullah2010}
Amanullah, R., Lidman, C., Rubin, D., {et~al.} 2010, APJ, 716, 712

\bibitem[{Amendola(2000)}]{Amendola2000}
Amendola, L. 2000, Phys. Rev. D, 62, 043511

\bibitem[{Baldwin(1977)}]{Baldwin1977}
Baldwin, J.~A. 1977, APJ, 214, 679

\bibitem[{Baskin \& Laor(2004)}]{Baskin2004}
Baskin, A. \& Laor, A. 2004, MNRAS, 350, L31

\bibitem[{Berk {et~al.}(2001)Berk, Richards, Bauer, Strauss, Schneider,
  Heckman, York, Hall, Fan, Knapp, {et~al.}}]{Berk2001}
Berk, D. E.~V., Richards, G.~T., Bauer, A., {et~al.} 2001, APJ, 122, 549

\bibitem[{Betoule {et~al.}(2014)Betoule, Kessler, Guy, Mosher, Hardin, Biswas,
  Astier, El-Hage, Konig, Kuhlmann, {et~al.}}]{Betoule2014}
Betoule, M., Kessler, R., Guy, J., {et~al.} 2014, A\&A, 568, A22

\bibitem[{Bian {et~al.}(2012)Bian, Fang, Huang, \& Wang}]{Bian2012}
Bian, W.-H., Fang, L.-L., Huang, K.-L., \& Wang, J.-M. 2012, MNRAS, 427, 2881

\bibitem[{Chang {et~al.}(2021)Chang, Xie, Liu, Ho, Dong, Han, \&
  Wang}]{Chang2021}
Chang, N., Xie, F., Liu, X., {et~al.} 2021, MNRAS, 503, 1987

\bibitem[{Clarkson \& Zunckel(2010)}]{Clarkson2010}
Clarkson, C. \& Zunckel, C. 2010, Phys. Rev. Lett., 104, 211301

\bibitem[{Condon {et~al.}(1998)Condon, Cotton, Greisen, Yin, Perley, Taylor, \&
  Broderick}]{Condon1998}
Condon, J.~J., Cotton, W., Greisen, E., {et~al.} 1998, APJ, 115, 1693

\bibitem[{Conley {et~al.}(2010)Conley, Guy, Sullivan, Regnault, Astier,
  Balland, Basa, Carlberg, Fouchez, Hardin, {et~al.}}]{Conley2010}
Conley, A., Guy, J., Sullivan, M., {et~al.} 2010, APJS, 192, 1

\bibitem[{Crittenden {et~al.}(2009)Crittenden, Pogosian, \&
  Zhao}]{Crittenden2009}
Crittenden, R.~G., Pogosian, L., \& Zhao, G.~B. 2009, JCAP, 2009, 025

\bibitem[{Croom {et~al.}(2002)Croom, Rhook, Corbett, Boyle, Netzer, Loaring,
  Miller, Outram, Shanks, \& Smith}]{Croom2002}
Croom, S., Rhook, K., Corbett, E., {et~al.} 2002, MNRAS, 337, 275

\bibitem[{Dietrich {et~al.}(2002)Dietrich, Hamann, Shields, Constantin,
  Vestergaard, Chaffee, Foltz, \& Junkkarinen}]{Dietrich2002}
Dietrich, M., Hamann, F., Shields, J., {et~al.} 2002, APJ, 581, 912

\bibitem[{Dong {et~al.}(2009)Dong, Wang, Wang, Fan, Wang, Zhou, \&
  Yuan}]{Dong2009}
Dong, X.-B., Wang, T.-G., Wang, J.-G., {et~al.} 2009, APJ, 703, L1

\bibitem[{Dultzin {et~al.}(2020)Dultzin, Marziani, De~Diego, Negrete, Del~Olmo,
  Mart{\'\i}nez-Aldama, D'Onofrio, Bon, Bon, \& Stirpe}]{Dultzin2020}
Dultzin, D., Marziani, P., De~Diego, J., {et~al.} 2020, Frontiers in Astronomy
  and Space Sciences, 6, 80

\bibitem[{Gao {et~al.}(2017)Gao, Wang, Shan, Li, \& Wang}]{Gao2017}
Gao, Z.~F., Wang, N., Shan, H., Li, X.~D., \& Wang, W. 2017, APJ, 849, 19

\bibitem[{Ge {et~al.}(2016)Ge, Bian, Jiang, Liu, \& Wang}]{Ge2016}
Ge, X., Bian, W.~H., Jiang, X.~L., Liu, W.~S., \& Wang, X.-F. 2016, MNRAS, 462,
  966

\bibitem[{Gunn {et~al.}(2006)Gunn, Siegmund, Mannery, Owen, Hull, Leger, Carey,
  Knapp, York, Boroski, {et~al.}}]{Gunn2006}
Gunn, J.~E., Siegmund, W.~A., Mannery, E.~J., {et~al.} 2006, AJ, 131, 2332

\bibitem[{Helfand {et~al.}(2015)Helfand, White, \& Becker}]{Helfand2015}
Helfand, D.~J., White, R.~L., \& Becker, R.~H. 2015, APJ, 801, 26

\bibitem[{Holsclaw {et~al.}(2010)Holsclaw, Alam, Sanso, Lee, Heitmann, Habib,
  \& Higdon}]{Holsclaw2010}
Holsclaw, T., Alam, U., Sanso, B., {et~al.} 2010, Phys. Rev. Lett., 105, 241302

\bibitem[{Hu \& Dodelson(2002)}]{Hu2002}
Hu, W. \& Dodelson, S. 2002, ARA\&A, 40, 171

\bibitem[{Huang \& Chang(2022)}]{Huang2022}
Huang, L. \& Chang, Z. 2022, MNRAS, 515, 1358

\bibitem[{Huterer \& Starkman(2003)}]{Huterer2003}
Huterer, D. \& Starkman, G. 2003, Phys. Rev. Lett., 90, 031301

\bibitem[{Kellermann {et~al.}(1994)Kellermann, Sramek, Schmidt, Green, \&
  Shaffer}]{Kellermann1994}
Kellermann, K., Sramek, R., Schmidt, M., Green, R., \& Shaffer, D. 1994, AJ,
  108, 1163

\bibitem[{Kellermann {et~al.}(1989)Kellermann, Sramek, Schmidt, Shaffer, \&
  Green}]{Kellermann1989}
Kellermann, K., Sramek, R., Schmidt, M., Shaffer, D., \& Green, R. 1989, AJ,
  98, 1195

\bibitem[{Kim(2011)}]{Kim2011}
Kim, A.~G. 2011, PASP, 123, 230

\bibitem[{Kinney {et~al.}(1990)Kinney, Rivolo, \& Koratkar}]{Kinney1990}
Kinney, A., Rivolo, A., \& Koratkar, A. 1990, APJ, 357, 338

\bibitem[{Kondo {et~al.}(1989)Kondo, Boggess, \& Maran}]{Kondo1989}
Kondo, Y., Boggess, A., \& Maran, S.~P. 1989, ARA\&A, 27, 397

\bibitem[{La~Franca {et~al.}(2014)La~Franca, Bianchi, Ponti, Branchini, \&
  Matt}]{La2014}
La~Franca, F., Bianchi, S., Ponti, G., Branchini, E., \& Matt, G. 2014, APJL,
  787, L12

\bibitem[{Li(2004)}]{Li2004}
Li, M. 2004, Phys. Lett. B, 603, 1

\bibitem[{Linder(2003)}]{Linder2003}
Linder, E.~V. 2003, Phys. Rev. Lett., 90, 091301

\bibitem[{Lyke {et~al.}(2020)Lyke, Higley, McLane, Schurhammer, Myers, Ross,
  Dawson, Chabanier, Martini, Des~Bourboux, {et~al.}}]{Lyke2020}
Lyke, B.~W., Higley, A.~N., McLane, J., {et~al.} 2020, APJS, 250, 8

\bibitem[{Maor {et~al.}(2002)Maor, Brustein, McMahon, \& Steinhardt}]{Maor2002}
Maor, I., Brustein, R., McMahon, J., \& Steinhardt, P.~J. 2002, Phys. Rev. D,
  65, 123003

\bibitem[{Mart{\'\i}nez-Aldama {et~al.}(2019)Mart{\'\i}nez-Aldama, Czerny,
  Kawka, Karas, Panda, Zaja{\v{c}}ek, \& {\.Z}ycki}]{Martinez2019}
Mart{\'\i}nez-Aldama, M.~L., Czerny, B., Kawka, D., {et~al.} 2019, APJ, 883,
  170

\bibitem[{Maziashvili(2007)}]{Maziashvili2007}
Maziashvili, M. 2007, Int. J. Mod. Phys. D, 16, 1531

\bibitem[{McLure \& Dunlop(2004)}]{Mclure2004}
McLure, R.~J. \& Dunlop, J.~S. 2004, MNRAS, 352, 1390

\bibitem[{Naidu {et~al.}(2022)Naidu, Oesch, van Dokkum, Nelson, Suess,
  Whitaker, Allen, Bezanson, Bouwens, Brammer, {et~al.}}]{Naidu2022}
Naidu, R.~P., Oesch, P.~A., van Dokkum, P., {et~al.} 2022,
  \href{https://arxiv.org/abs/2207.09434}{arXiv:2207.09434}

\bibitem[{Netzer {et~al.}(1992)Netzer, Laor, \& Gondhalekar}]{Netzer1992}
Netzer, H., Laor, A., \& Gondhalekar, P. 1992, MNRAS, 254, 15

\bibitem[{Niko{\l}ajuk \& Walter(2012)}]{Nikolajuk2012}
Niko{\l}ajuk, M. \& Walter, R. 2012, MNRAS, 420, 2518

\bibitem[{P{\^a}ris {et~al.}(2012)P{\^a}ris, Petitjean, Aubourg, Bailey, Ross,
  Myers, Strauss, Anderson, Arnau, Bautista, {et~al.}}]{Paris2012}
P{\^a}ris, I., Petitjean, P., Aubourg, {\'E}., {et~al.} 2012, A \& A, 548, A66

\bibitem[{P{\^a}ris {et~al.}(2011)P{\^a}ris, Petitjean, Rollinde, Aubourg,
  Charlassier, Delubac, Hamilton, Le~Goff, Palanque-Delabrouille, Peirani,
  {et~al.}}]{Paris2011}
P{\^a}ris, I., Petitjean, P., Rollinde, E., {et~al.} 2011, A \& A, 530, A50

\bibitem[{P{\^a}ris {et~al.}(2017)P{\^a}ris, Petitjean, Ross, Myers, Aubourg,
  Streblyanska, Bailey, Armengaud, Palanque-Delabrouille, Y{\`e}che,
  {et~al.}}]{Paris2017}
P{\^a}ris, I., Petitjean, P., Ross, N.~P., {et~al.} 2017, A\&A, 597, A79

\bibitem[{Peebles \& Ratra(2003)}]{Peebles2003}
Peebles, P. J.~E. \& Ratra, B. 2003, Rev. Mod. Phys, 75, 559

\bibitem[{Ratra \& Peebles(1988)}]{Ratra1988}
Ratra, B. \& Peebles, P.~J. 1988, Phys. Rev. D, 37, 3406

\bibitem[{Reid {et~al.}(2019)Reid, Pesce, \& Riess}]{Reid2019}
Reid, M., Pesce, D.~W., \& Riess, A. 2019, APJL, 886, L27

\bibitem[{Riess {et~al.}(1998)Riess, Filippenko, Challis, Clocchiatti, Diercks,
  Garnavich, Gilliland, Hogan, Jha, Kirshner, {et~al.}}]{Riess1998}
Riess, A.~G., Filippenko, A.~V., Challis, P., {et~al.} 1998, AJ, 116, 1009

\bibitem[{Risaliti \& Lusso(2015)}]{Risaliti2015}
Risaliti, G. \& Lusso, E. 2015, APJ, 815, 33

\bibitem[{Sahni \& Starobinsky(2006)}]{Sahni2006}
Sahni, V. \& Starobinsky, A. 2006, INT J MOD PHYS D, 15, 2105

\bibitem[{Sako {et~al.}(2018)Sako, Bassett, Becker, Brown, Campbell, Wolf,
  Cinabro, D’andrea, Dawson, DeJongh, {et~al.}}]{Sako2018}
Sako, M., Bassett, B., Becker, A.~C., {et~al.} 2018, Publ. Astron. Soc. Aust.,
  130, 064002

\bibitem[{Scolnic {et~al.}(2018)Scolnic, Jones, Rest, Pan, Chornock, Foley,
  Huber, Kessler, Narayan, Riess, {et~al.}}]{Scolnic2018}
Scolnic, D.~M., Jones, D., Rest, A., {et~al.} 2018, APJ, 859, 101

\bibitem[{Seikel {et~al.}(2012)Seikel, Clarkson, \& Smith}]{Seikel2012}
Seikel, M., Clarkson, C., \& Smith, M. 2012, J. Cosmol. Astropart. Phys., 2012,
  036

\bibitem[{Shemmer \& Lieber(2015)}]{Shemmer2015}
Shemmer, O. \& Lieber, S. 2015, APJ, 805, 124

\bibitem[{Shen {et~al.}(2008)Shen, Greene, Strauss, Richards, \&
  Schneider}]{Shen2008}
Shen, Y., Greene, J.~E., Strauss, M.~A., Richards, G.~T., \& Schneider, D.~P.
  2008, APJ, 680, 169

\bibitem[{Shen \& Ho(2014)}]{Shen2014}
Shen, Y. \& Ho, L.~C. 2014, Nature, 513, 210

\bibitem[{Shen \& M{\'e}nard(2012)}]{Shen2012}
Shen, Y. \& M{\'e}nard, B. 2012, APJ, 748, 131

\bibitem[{Shen {et~al.}(2011)Shen, Richards, Strauss, Hall, Schneider, Snedden,
  Bizyaev, Brewington, Malanushenko, Malanushenko, {et~al.}}]{Shen2011}
Shen, Y., Richards, G.~T., Strauss, M.~A., {et~al.} 2011, APJS, 194, 45

\bibitem[{Shields(2006)}]{Shields2006}
Shields, J.~C. 2006, arXiv preprint astro-ph/0612613

\bibitem[{Stocke {et~al.}(1992)Stocke, Morris, Weymann, \& Foltz}]{Stocke1992}
Stocke, J.~T., Morris, S.~L., Weymann, R.~J., \& Foltz, C.~B. 1992, APJ, 396,
  487

\bibitem[{Strittmatter {et~al.}(1980)Strittmatter, Hill, Pauliny-Toth, Steppe,
  \& Witzel}]{Strittmatter1980}
Strittmatter, P., Hill, P., Pauliny-Toth, I., Steppe, H., \& Witzel, A. 1980,
  A\&A, 88, L12

\bibitem[{Sulentic {et~al.}(2000)Sulentic, Marziani, \&
  Dultzin-Hacyan}]{Sulentic2000}
Sulentic, J., Marziani, P., \& Dultzin-Hacyan, D. 2000, ARA\&A, 38, 521

\bibitem[{Sulentic {et~al.}(2007)Sulentic, Bachev, Marziani, Negrete, \&
  Dultzin}]{Sulentic2007}
Sulentic, J.~W., Bachev, R., Marziani, P., Negrete, C.~A., \& Dultzin, D. 2007,
  APJ, 666, 757

\bibitem[{Tacconi {et~al.}(2018)Tacconi, Genzel, Saintonge, Combes,
  Garc{\'\i}a-Burillo, Neri, Bolatto, Contini, Schreiber, Lilly,
  {et~al.}}]{Tacconi2018}
Tacconi, L.~J., Genzel, R., Saintonge, A., {et~al.} 2018, APJ, 853, 179

\bibitem[{Urry \& Padovani(1995)}]{Urry1995}
Urry, C.~M. \& Padovani, P. 1995, Publ. Astron. Soc. Aust., 107, 803

\bibitem[{Vestergaard \& Peterson(2006)}]{Vestergaard2006}
Vestergaard, M. \& Peterson, B.~M. 2006, APJ, 641, 689

\bibitem[{Xu {et~al.}(2008)Xu, Bian, Yuan, \& Huang}]{Xu2008}
Xu, Y., Bian, W.~H., Yuan, Q.-R., \& Huang, K.~L. 2008, MNRAS, 389, 1703

\bibitem[{Zhao {et~al.}(2012)Zhao, Crittenden, Pogosian, \& Zhang}]{Zhao2012}
Zhao, G.-B., Crittenden, R.~G., Pogosian, L., \& Zhang, X. 2012, Phys. Rev.
  Lett., 109, 171301

\bibitem[{Zheng {et~al.}(1995)Zheng, Kriss, \& Davidsen}]{Zheng1995}
Zheng, W., Kriss, G.~A., \& Davidsen, A.~F. 1995, APJ, 440, 606

\end{thebibliography}

\end{document}